\documentstyle[aps,eqsecnum,epsf]{revtex} 
\draft

\begin{document}
\tighten
\title{Generation of Scalar-Tensor Gravity Effects in Equilibrium State 
Boson Stars\footnote{{\tt gr-qc/9708071},  To appear in {\it Classical and
Quantum Gravity}.}}
\author{G. L. Comer\footnote{Email: {\tt comergl@pxa.slu.edu}}}
\address{Department of Science and Mathematics, 
Parks College of Engineering and Aviation\\
Saint Louis University, 
P.O. Box 56907, 
St. Louis, MO 63156-0907, USA}
\author{Hisa-aki Shinkai\footnote{Email: {\tt shinkai@wurel.wustl.edu}}}
\address{Department of Physics, Washington University, 
St. Louis, MO 63130-4899, USA}
\date{September 25, 1997 (revised version)}
\maketitle
\begin{abstract}
Boson stars in zero-, one-, and two-node equilibrium states are modeled 
numerically within the framework of Scalar-Tensor Gravity.  The 
complex scalar field is taken to be both massive and self-interacting.  
Configurations are formed in the case of a linear gravitational scalar 
coupling (the Brans-Dicke case) and a quadratic coupling which has been 
used previously in a cosmological context.  The coupling parameters and 
asymptotic value for the gravitational scalar field are chosen so that 
the known observational constraints on Scalar-Tensor Gravity are 
satisfied.  It is found that the constraints are so restrictive that 
the field equations of General Relativity and Scalar-Tensor Gravity 
yield virtually identical solutions.  We then use catastrophe theory 
to determine the dynamically stable configurations.  It is found that 
the maximum mass allowed for a stable state in Scalar-Tensor Gravity 
in the present cosmological era is essentially unchanged from that 
of General Relativity.  We also construct boson star configurations 
appropriate to earlier cosmological eras and find that the maximum 
mass for stable states is smaller than that predicted by General 
Relativity, and the more so for earlier eras.  However, our results 
also show that if the cosmological era is early enough then only states 
with positive binding energy can be constructed.            
\end{abstract}
\pacs{PACS number(s): 04.40.-b, 04.40.Dg, 04.50.+h}
\section{Introduction} \label{sec1}
General Relativity describes very well the gravitational interaction.  
However, it still has competitors, one of which is Scalar-Tensor 
Gravity (see Will \cite{CW} for a review).  Besides the metric, 
Scalar-Tensor Gravity has an additional scalar field.  Motivation for 
this additional field can be found in Dicke's \cite{D} discussions about 
arbitrariness in the measurement of lengths, times, and masses.  There 
is also the fact that dilaton gravity is the low string tension limit 
of string theory (see, for instance, Damour and Polyakov \cite{DP}).  
Motivation for the theory itself, is the attractor mechanism discussed 
by Damour and Nordtvedt \cite{ND} (see also \cite{GQ,BM}).  They 
demonstrate that the homogeneous and isotropic expansion of the universe 
forces a series of Scalar-Tensor cosmological solutions to be virtually 
indistinguishable from General Relativity solutions at late cosmological 
times.  This implies that today the weak-field differences between 
Scalar-Tensor Gravity and General Relativity are small.  It is the 
purpose here to see if significant strong-field differences between the 
two can be generated for equilibrium state boson stars.

A boson star consists of massive scalar particles (i.e., excitations 
of a complex scalar field) that are held together by gravity, a 
self-interaction, or both.  They were first discussed by Kaup \cite{K} 
and then by Ruffini and Bonazzola \cite{RB} (for a thorough review, 
see Liddle and Madsen \cite{LM}).  They can form stable configurations 
having negative binding energy.  If we include a self-interaction, then 
their maximum allowed stable mass comes to the order of a solar mass 
\cite{CSW}.  It is even speculated that they are a form of dark matter
and that they can be created during a phase transition in the early 
universe (see Frieman et al \cite{FGGK}).  Unlike boson stars formed 
today, those formed in the early universe would have done so in an 
era when even the weak-field differences between Scalar-Tensor Gravity
and General Relativity are significant.  

In terms of their equilibrium characteristics, there are infinitely 
many different boson star configurations for a given central density, 
depending on how many nodes, or zeroes, there are in the matter scalar 
field.  Furthermore, the ground state, or zero-node configuration, 
has the lowest ADM mass, the higher-node configurations then have 
successfully higher masses.  With this in mind, there are three 
different ways that we try to generate non-trivial Scalar-Tensor 
effects: (i) build zero- and higher-node configurations for three 
different values of the matter scalar field self-interaction, (ii) 
build sequences of zero-node solutions, where different sequences have 
different values for the self-interaction, and (iii) repeat (i) and (ii) 
for two different Scalar-Tensor theories.   

The two varieties of Scalar-Tensor Gravity considered here use the 
linear Brans-Dicke coupling \cite{BD} and the quadratic coupling used by 
Damour and Nordtvedt \cite{ND}.  Experiments performed in our solar system 
have placed a lower bound on the slope of the Brans-Dicke linear coupling 
\cite{CW}.  We will only consider Brans-Dicke boson stars that are 
consistent with this lower bound.  As for the quadratic coupling, we 
impose boundary conditions such that the gravitational scalar can have 
an asymptotic value equal to the ambient, cosmological value existing at 
the present time, as well as earlier times.  The ambient value is derived 
from a cosmological solution that exhibits the attractor mechanism of 
Damour and Nordtvedt \cite{ND}.  When the present-day ambient value is 
used, then the present-day observational constraints are satisfied.
 
Boson stars in Scalar-Tensor Gravity have been investigated previously 
by Gundersen and Jensen \cite{GJ}, who used the Brans-Dicke \cite{BD} 
coupling, and by Torres \cite{torres}, who looked at three different 
classes of couplings.  Both discussions considered only zero-node 
configurations.  Here we will consider solutions containing up to two 
nodes.  Furthermore, neither of the previous discussions considered 
the question of stability. Here, we will apply catastrophe theory to the 
binding energy to determine the change of stability for the zero-node 
solutions.  Finally, we, like Torres, will look at boson stars formed in 
earlier eras of the history of the universe.   

We will show below that stable, zero-node boson star solutions in 
Scalar-Tensor Gravity are, for all practical purposes, indistinguishable 
from those of General Relativity in the present cosmological era.  We 
also show that the same holds for the higher-node solutions, regardless 
of the strength of the self-interaction.   It is for this reason that we 
construct zero-node boson star configurations appropriate to earlier 
cosmological eras, so as to determine if the indistinquishability of the 
solutions is a generic feature of boson stars or a consequence of the 
cosmological conditions.  We shall see that it is the cosmology that is 
predominate, in agreement with the suggestion of Torres \cite{torres} 
that masses can decrease from their General Relativity values for earlier 
cosmological eras.  But moreover, we use  catastrophe theory to show that 
the maximum, stable mass decreases as the cosmological time decreases.  
However, when the cosmological era is early enough, we find that only 
positive binding energies result so that no stable configuration can be 
constructed.  

The fact that negative binding energy states can exist at all is a priori 
a non-trivial result.   Such states can be realized within General 
Relativity \cite{WT} because a boson star's typical mass is $M \sim 
1/G m$ and its total number of particles is $N_p \sim 1/G m^2$---where 
$G$ is the Newton gravitational coupling and $m$ is the individual 
mass of the scalar particles making up the star---and so both are of 
comparable value in the binding energy $M - m N_p$.  However, one way 
in which Scalar-Tensor Gravity differs from General Relativity is in 
how it locally varies the Newton gravitational coupling (see, for 
instance, Bekenstein \cite{B} and Bekenstein and Meisels \cite{BM}, 
and references therein).  The inherent non-linear features of the 
Scalar-Tensor field equations suggest that a spatially varying G, 
which at a very basic level controls the strength of the gravitational 
interaction between two particles, should lead to qualitative 
differences in binding energy between General Relativity and 
Scalar-Tensor Gravity.  We shall derive below a necessary condition 
for negative binding energies to exist.  

In the next section we will introduce the field equations for the 
gravitational fields and the matter.  In Sec. \ref{sec3} we write 
down the equations in spherically symmetric form.  Similar to the 
approach of Friedberg et al \cite{FLP}, we write down a total energy 
functional from which the field equations can be derived.  It is 
also here that we discuss the boundary conditions and the numerical 
technique.  In Sec. \ref{sec4} we give a detailed account of the 
solutions containing up to two nodes for a typical central value of 
the matter scalar field.  In Sec. \ref{sec5}, we investigate the 
stability of the ground state configurations through the use of 
catastrophe theory.  In Sec. \ref{sec6} we discuss boson stars formed 
in earlier cosmological eras.  Finally, in Sec. \ref{sec7} we make some 
concluding remarks.  Units such that $c = 1$ and $\hbar = 1$ will be 
used.  This implies that the scalar field mass $m$ is an inverse length 
(actually, the inverse Compton wavelength of the scalar particles) and 
the bare gravitational constant $G_{*}$ has units of length squared.  

\section{The field equations}\label{sec2}

The action for our system of Scalar-Tensor Gravity coupled to a 
self-interacting, complex scalar field in the physical, ``Jordan-frame'' 
is
\begin{eqnarray}
    S &=& {1\over 16 \pi} \int d^4x  \sqrt{-\tilde{g}} 
          \left[\phi \tilde{R} -\phi^{-1} \omega(\phi) 
          \tilde{g}^{\mu\nu}  \partial_\mu \phi \partial_\nu \phi
          \right] - \nonumber \\
          &&\int {\rm d}^4x \sqrt{- \tilde{g}} 
          \left[{1 \over 2} \tilde{g}^{\mu \nu} \partial_{\mu} 
          \psi^{\dag} \partial_{\nu} \psi + \left({m^2 \over 2} 
          \psi^{\dag} \psi + V\left(\psi^{\dag} \psi\right)\right)
          \right] + B.T. \ , \label{2.1} 
\end{eqnarray}
where $B.T.$ represents the boundary terms that can be added to 
subtract out the second-order derivatives coming from $R$ \cite{JWY,GH} 
(the explicit form being identical to that used by Friedberg et al 
\cite{FLP} and will be given below in the ``Einstein-frame'').  The 
gravitational scalar is $\phi$ and $\omega(\phi)$ is the ``Jordan-frame'' 
coupling of $\phi$ to the matter.  The complex scalar $\psi$ (with its 
complex conjugate being $\psi^{\dagger}$) has mass $m$ and is 
self-interacting through the potential $V\left(\psi^{\dag} \psi\right)$.
 
There is an alternative representation of the action above, for the 
so-called ``Einstein-frame.''  The transition to this frame is effected 
by the conformal transformation 
\begin{equation}
   \tilde{g}_{\mu \nu} =  e^{2 a(\varphi)} g_{\mu \nu} \ , \label{2.2}
\end{equation}
where
\begin{equation}
    \phi^{- 1} = G_* e^{2 a(\varphi)} \label{2.3}
\end{equation}
and  $a(\varphi)$ is the functional transformation from 
$\phi$ to the 
``Einstein-frame'' gravitational scalar $\varphi$.  The relationship 
between $\omega(\phi)$ and $a(\varphi)$ is obtained from
\begin{equation}
    \alpha^2 = (2 \omega + 3)^{- 1} \ , \label{2.4}
\end{equation}
where
\begin{equation}
   \alpha(\varphi) \equiv {\partial a \over \partial \varphi} \ . 
        \label{2.5}
\end{equation}
The action in the ``Einstein-frame'' is thus
\begin{eqnarray}
  S &=& {1 \over 16 \pi G_{*}} \int {\rm d}^4x \sqrt{-g} \left[R - 
      2 g^{\mu \nu} \partial_{\mu} \varphi \partial_{\nu} \varphi 
      \right]\nonumber \\
      &&- \int {\rm d}^4x \sqrt{-g} \left[{1 \over 2} 
      e^{2 a(\varphi)} g^{\mu \nu} \partial_{\mu} \psi^{\dag} 
      \partial_{\nu} \psi + e^{4 a(\varphi)} \left({m^2 \over 2} 
      \psi^{\dag} \psi + V\left(\psi^{\dag} \psi\right)\right)\right] 
      + B.T. \ . \label{2.6}
\end{eqnarray}
It does not deliver exactly General Relativity because the 
metric $g_{\mu \nu}$ is not the true, physical metric that encodes the 
distance between spacetime points.  However, the ``Einstein-frame'' 
does deliver equations that are similar enough to General Relativity 
that we will use it for our calculations. 

The ``Einstein-frame'' stress-energy tensor is
\begin{equation}
   T_{\mu \nu} = {1 \over 2} e^{2 a(\varphi)} \left(\partial_{\mu} 
                \psi^{\dag} \partial_{\nu} \psi + \partial_{\nu} 
                \psi^{\dag} \partial_{\mu} \psi\right) - {1 \over 2} 
                e^{2 a(\varphi)} \left( \partial_{\tau} \psi^{\dag} 
                \partial^{\tau} \psi + e^{2 a(\varphi)} \left[m^2 
                \psi^{\dag} \psi + 2 V(\psi^{\dag} \psi)\right]\right) 
                g_{\mu \nu} \ . \label{2.7}
\end{equation}
The gravitational field equations for $g_{\mu \nu}$ and $\varphi$ are
\begin{equation}
   G_{\mu \nu} = 8 \pi G_{*} T_{\mu \nu} + 2 \partial_{\mu} \varphi 
               \partial_{\nu} \varphi -  \partial_{\tau} \varphi 
               \partial^{\tau} \varphi g_{\mu \nu} \label{2.8}
\end{equation}
and
\begin{equation}
   \nabla_{\sigma} \nabla^{\sigma} \varphi = - 4 \pi \alpha T \ , 
               \label{2.9}
\end{equation}
where $T$ is the trace of the stress-energy tensor.  The matter field 
equations are
\begin{equation}
   \nabla_{\sigma} \nabla^{\sigma} \psi^{\dag} + 2 \alpha 
                \partial_{\tau} \psi^{\dag} \partial^{\tau} \varphi 
                = e^{2 a(\varphi)} \left(m^2 \psi^{\dag} + 2 {\partial 
                V \over \partial \psi}\right) \label{2.10a}
\end{equation}
and
\begin{equation}
   \nabla_{\sigma} \nabla^{\sigma} \psi + 2 \alpha  \partial_{\tau} 
                \psi \partial^{\tau} \varphi = e^{2 a(\varphi)} 
                \left(m^2 \psi + 2 {\partial V \over \partial 
                \psi^{\dag}}\right) \ . \label{2.10b}
\end{equation}

The coupling function $a(\varphi)$ is a priori unknown. There are, 
however, some theoretical reasons to motivate explicit forms.  
Furthermore, once a particular form for $a(\varphi)$ is taken, there 
are experimental constraints \cite{CW,DEF} that can be imposed (using 
Eqs. (\ref{3b.2}) and (\ref{3b.3}) below).  The two forms for 
$a(\varphi)$ that will be used here are the Brans-Dicke coupling
\begin{equation}
    a(\varphi) = {\varphi - \varphi_{\infty} \over \sqrt{2 \omega_{BD} 
                 + 3}} \label{2.11}
\end{equation} 
and the quadratic coupling
\begin{equation}
    a(\varphi) = {1 \over 2} \kappa \left(\varphi^2 - \varphi^2_{\infty}
                 \right) \ , \label{2.12}
\end{equation}
which was the particular form considered by Damour and Nordtvedt 
\cite{ND} (except for the additive constant).  The term 
$\varphi_{\infty}$ represents the asymptotic value of the gravitational 
scalar field.  It is known from Solar System observations that 
$\omega_{BD} > 500$; likewise, it is known from observations of binary 
pulsars that $\kappa > - 5$ \cite{DEF}.  

As for the complex scalar field, we will be considering a 
self-interaction of the form
\begin{equation}
    V\left(\psi^{\dag} \psi\right) = {\Lambda \over 4} 
                 \left(\psi^{\dag} \psi\right)^2 \ . \label{2.13}
\end{equation}
The strength of the self-interaction, $\Lambda$, will be taken to be 
positive.  Because the potential $V\left(\psi^{\dag} \psi\right)$ is 
a functional of $\psi^{\dag} \psi$ then it preserves the global U(1) 
gauge symmetry ($\psi \to e^{i \sigma} \psi$, where $\sigma$ is a 
constant) present in the theory.  This symmetry results in a conserved 
current, whose explicit form in the ``Jordan-frame'' is
\begin{equation}
    \tilde{J}^{\mu} = {i \over 2} e^{-2 a(\varphi)} g^{\mu \nu} 
              \left(\psi \partial_{\nu} \psi^{\dag} - \psi^{\dag} 
              \partial_{\nu} \psi\right) \ . \label{2.14}
\end{equation}
This conserved current leads to a conserved charge, which is $N_p$, 
the number of particles making up the star (cf., Eq. (\ref{3a.8}) 
below).  

\section{The equilibrium state equations} \label{sec3}

\subsection{The equilibrium state field equations} \label{sec3a}

The spacetimes considered here are spherically symmetric and static, 
with the ``Einstein-frame'' metric taking the form
\begin{equation}
     {\rm d}s^2 = - N^2(r) {\rm d}t^2 + A(r){\rm d}r^2 + r^2 
                  \left[{\rm d}\theta^2 + {\rm sin}^2 \theta {\rm d}
                  \phi^2\right] \ . \label{3a.1}
\end{equation}
The gravitational scalar, which is real, is assumed also to be spherically 
symmetric and static:
\begin{equation}
     \varphi = \varphi(r) \ . \label{3a.2}
\end{equation}
As for the matter scalar field, Friedberg et al \cite{FLP} show that 
the minimum energy configurations are those for which 
\begin{equation}
     \psi = e^{- i \Omega t} \Phi(r) \ , \label{3a.3}
\end{equation}
where $\Omega$ is real and positive.  Their proof (see the Appendix 
in \cite{FLP}) also goes through for scalar-tensor gravity  (making use 
of an energy functional defined as below in Eq. (\ref{3a.7})), and so we 
will take $\psi$ to have this form. 

To write down the field equations for this system we could take Eqs. 
(\ref{3a.1}-\ref{3a.3}) and insert them into the full field equations 
written earlier.  However, it is more useful to insert Eqs. 
(\ref{3a.1}-\ref{3a.3}) directly into Eq. (\ref{2.6}), and then derive 
the field equations from the reduced action.  Let $L_{matter}$ represent 
the reduced ``matter'' Lagrangian (which includes the contribution 
due to $\varphi$), $L_{grav}$ the reduced gravitational Lagrangian, and 
$L = L_{matter} + L_{grav}$ the total Lagrangian, i.e., $S = \int 
{\rm d}t L$ where $S$ is the total reduced action.  Then for our 
particular system we find  
\begin{equation}
    L_{matter} = - {m^2 \over G_{*}} \int_0^{\infty} {\rm d}r~r^2 N 
           \sqrt{A} (u + v - w) \label{3a.4a}
\end{equation} 
and
\begin{equation}
    L_{grav} = {1 \over 2G_{*}} \int_0^{\infty} {\rm d}r~N \left(
           \sqrt{A} - 2 \left[1 + r {N^{\prime} \over N}\right] + 
           {1 \over \sqrt{A}} \left[1 + 2 r {N^{\prime} \over N}\right]
           \right) \ , \label{3a.4b}
\end{equation}    
where
\begin{equation}
    w = {2 \pi G_{*} \Omega^2 \over m^2 N^2} e^{2 a(\varphi)} 
        \Phi^2 \ , \label{3a.5a}
\end{equation}
\begin{equation}
    u = 2 \pi G_{*} \left(1 + {\Lambda \over 2 m^2} \Phi^2\right) e^{4 
        a(\varphi)} \Phi^2 \ , \label{3a.5b}
\end{equation}
and
\begin{equation}
    v = {1 \over 2 m^2 A} \left(4 \pi G_{*} e^{2 a(\varphi)} 
        \left[\Phi^{\prime}\right]^2 + \left[\varphi^{\prime}\right]^2
        \right) \ . \label{3a.5c}
\end{equation}
Note that the gravitational Lagrangian does {\it not} follow just 
from the Hilbert action, it also includes the contributions due to 
the boundary term $B.T.$ of Eq. (\ref{2.6}):
\begin{equation}
    B.T. = {1 \over 4 G_{*}} \left.\left(\int {\rm d}t \left[{2 N r^2 
           \over \sqrt{A}}\left({N^{\prime} \over N} + {2 \over r}
           \right) - 4 N r\right]\right)\right|_{r = 0}^{r = \infty}\ . 
           \label{3a.6}
\end{equation} 

We can obtain the total energy functional $E$ (i.e., Hamiltonian) from 
the Lagrangian via the standard canonical transformation that replaces 
$\dot{\psi}$ with its conjugate momentum:
\begin{eqnarray}
    E &=& \Omega N_p - L \nonumber \\
      &=& {m^2 \over G_{*}} \int_0^{\infty} {\rm d}r~r^2 N \sqrt{A} 
          \left(w + u + v\right) - L_{grav} \ , \label{3a.7}
\end{eqnarray}
where $N_p$ is the conserved total particle number (in the physical 
``Jordan-frame'') and is given by
\begin{equation}
    N_p = {2 m^2 \over G_{*} \Omega} \int_0^{\infty} {\rm d}r~r^2 
          N \sqrt{A} w \ . \label{3a.8}
\end{equation}
By construction it is the case that 
\begin{equation}
   \Omega = \left.\left[{\partial E \over \partial N_p}\right]
        \right|_{\Phi,\varphi,A,N} = {{\rm d}M \over {\rm d}N_p} \ , 
        \label{3a.9}
\end{equation}
where $M$ is the ``on-shell'' total mass-energy (given below in Eq. 
(\ref{3c.1})), and $N_p$ is also ``on-shell.''

There are now two ways to derive the field equations: one can vary $L$ 
keeping $\Omega$ fixed to get them, or one can vary $E$ keeping $N_p$ 
fixed.  Either way gives the same results, which are
\begin{equation}
   {A^{\prime} \over A} = {1 - A \over r} + 2 m^2 r A \left(w + u + 
               v\right) \ , \label{3a.10a}
\end{equation}
\begin{equation}
   {2 N^{\prime} \over N} = - {1 - A \over r} + 2 m^2 r A \left(w - u 
               + v\right) \ , \label{3a.10b}
\end{equation}
\begin{equation}
   \left({r^2 N \varphi^{\prime} \over \sqrt{A}}\right)^{\prime} = 4 
        \pi G_{*} m^2 r^2 N \sqrt{A} e^{2 a(\varphi)} \alpha
        \left(\left[2 e^{2 a(\varphi)} - {\Omega^2 \over m^2 N^2}
        \right] \Phi^2 + {1 \over A m^2} \left[\Phi^{\prime}\right]^2 
        + {\Lambda \over m^2} e^{2 a(\varphi)} \Phi^4\right) 
        \ , \label{3a.10c}
\end{equation}
and
\begin{equation}
   \left({r^2 N e^{2 a(\varphi)} \Phi^{\prime} \over \sqrt{A}}
         \right)^{\prime} = m^2 r^2 N \sqrt{A} e^{2 a(\varphi)} 
         \left(e^{2 a(\varphi)} - {\Omega^2 \over m^2 N^{2}}\right) 
         \Phi + \lambda r^2 N \sqrt{A} e^{4 a(\varphi)} \Phi^3 
         \ . \label{3a.10d}
\end{equation}

\subsection{Boundary conditions} \label{sec3b}

The boundary conditions for this system of equations must take into 
account three things: (i) the solutions must be geometrically 
regular at the origin; (ii) the solutions must yield an asymptotically 
flat spacetime; and (iii) the solutions must take into account the 
cosmological input for both the coupling $a(\varphi)$ as well as 
$\varphi$.  

Geometrical regularity at the origin means there is no conical 
singularity, i.e., the proper radius divided by the proper circumference 
should reduce to $2 \pi$ at $r = 0$.  This implies that $A(0) = 1$.  
Also, to maintain regularity in the field equations as $r \to 0$, we 
impose that $\Phi^{\prime}(0) = 0$ and $\varphi^{\prime}(0) = 0$.    

For a purely technical reason (to be discussed below), we desire 
solutions that are asymptotically flat in both the Jordan and 
Einstein frames.  That is, we want both $\tilde{g}_{\mu \nu}$ and 
$g_{\mu \nu}$ to reduce to the flat spacetime metric at spatial 
infinity.  The implication of this is that the value of 
$\varphi_{\infty} \equiv \varphi(\infty)$ must be such that 
$a(\varphi_{\infty}) = 0$.  The other outcome is that both $N$ and 
$A$ approach one at spatial infinity.  

For the Brans-Dicke coupling, $\Phi_c \equiv \Phi(0)$ and 
$\varphi_{\infty}$ are the only freely specified field values.  The 
value of $N(0)$ is not specified freely, but rather is determined so 
that $\Phi_{\infty} \equiv \Phi(\infty) = 0$.  The value of $\varphi$ 
at the origin is not specified freely; it must be determined in such 
a way that the solution for $\varphi$ goes to $\varphi_{\infty}$ at 
spatial infinity.  We will use the freedom to add an arbitrary constant 
to the Brans-Dicke coupling $a(\varphi)$ (to be discussed in more detail 
below) so that all the solutions we consider have $\varphi_{\infty} = 0$.  

For the quadratic coupling, again $\Phi_c$ can be specified freely, but 
we use the cosmological model of Damour and Nordtvedt \cite{ND} to 
calculate $\varphi_{\infty}$, i.e.,
\begin{equation}
    \varphi_{\infty} \sim {\alpha_R \over \kappa} \left(1 - {3 
            \over 8 \kappa}\right)^{- 1/2} e^{-3 p/4} {\rm sin} \left({3 
            \over 4} \sqrt{{8 \kappa \over 3} - 1} p + {\rm arctan}
            \sqrt{{8 \kappa \over 3} - 1}\right) \ , \label{3b.1}
\end{equation}
where $\alpha_R \sim 1$ is the value of $\alpha$ for the universe 
at the end of the radiation dominated era and $p$ is a measure of the 
time since this era (i.e., today, $p \sim 10$); in terms of the 
cosmological redshift $z$ we have $z \sim e^{10-p} - 1$.  Note that 
we must take $\kappa \equiv \partial \alpha/\partial \varphi > 3/8$.  

Damour and Nordtvedt used a form of the scalar-tensor coupling without 
an additive constant (cf. Eq. (\ref{2.12})).  Fortunately, it can be 
shown that putting in such a constant does not change the solution above 
for $\varphi_{\infty}$.  Furthermore, the addition of such a constant 
will not affect the predictions of Damour and Nordvedt for the PPN 
(Parametrized Post-Newtonian) parameters $\gamma - 1$ and $\beta - 1$ or 
the time rate-of-change of the Newton coupling, since each of these only 
depend on the present day values for $\alpha$ and $\kappa$:
\begin{equation}
   \gamma - 1 = - {2 \alpha^2 \over 1 + \alpha^2} \label{3b.2}
\end{equation}
and
\begin{equation}
   \beta - 1 = - {1 \over 8} \kappa (\gamma + 1) (\gamma - 1) \ . 
               \label{3b.3} 
\end{equation}
Observational constraints (see Will \cite{CW}, and references therein) 
yield $|\gamma - 1| < 2 \times 10^{- 3}$ and $|\beta - 1| < 2 \times 
10^{- 3}$.

\subsection{ADM Mass} \label{sec3c}      

A consequence of determining $\varphi$ at spatial infinity such that 
$a(\varphi_{\infty}) = 0$ is that the ADM mass $M$ is given by 
\begin{equation}
    G_{*} M = m^2 \int_0^{\infty} {\rm d}r~r^2 \left(w + u + 
        v\right)\ . \label{3c.1}
\end{equation}
This can be shown as follows: the ``Jordan-frame'' ADM mass $M_J$ is 
given by
\begin{equation}
    G_{*} M_J = \lim_{r \to \infty} {r \over 2} \left(1 - 1/
              \tilde{g}_{r r}\right) \ . \label{3c.2}
\end{equation}
The similar ``Einstein-frame'' ADM mass $M_E$ is
\begin{equation}
    G_{*} M_E = \lim_{r \to \infty} {r \over 2} \left(1 - 1/g_{r r}
              \right) \ . \label{3c.3}
\end{equation}
However, since $\tilde{g}_{r r} = e^{2 a(\varphi)} g_{r r}$ and 
$a(\varphi_{\infty}) = 0$, then the limits on the right-hand-sides are 
equal and therefore $M_J = M_E \equiv M$.  Eq. (\ref{3c.1}) follows 
using the integrated form of Eq. (\ref{3a.10a}).  Furthermore, it can 
be shown that the value for $M$ one obtains from Eq. (\ref{3c.1}) is 
the same as that delivered by the ``on-shell'' value for the total 
energy functional $E$ of Eq. (\ref{3a.7}).

\subsection{Necessary condition for negative binding energy states} 
\label{sec3d}

If the condition 
\begin{equation}
     2 > \Omega/m \label{3d.1}
\end{equation}
is satisfied, then negative binding energy states can be constructed.  
This can be established by first adding together Eqs. (\ref{3a.10a}) 
and (\ref{3a.10b}), which results in 
\begin{equation}
    \left({\rm ln} \left[N^2 A\right]\right)^{\prime} = 4 m^2 r A 
            (v + w) \ . \label{3d.2}
\end{equation}
The right-hand-side of this equation is positive-definite so that 
$\left({\rm ln} \left[N^2 A\right]\right)^{\prime} > 0$ for all $r$.  
Our choice of coordinates and boundary conditions dictate that 
\begin{equation}
    \lim_{r \to \infty} {\rm ln} \left(N^2 A\right) = 0 \ . 
            \label{3d.3}
\end{equation}
Therefore, ${\rm ln} \left(N^2 A\right) < 0$ for all finite $r$ and 
thus $N^2 A < 1$.  Because the integrand of $M$ is positive definite 
(for positive $\Lambda$), and $N \sqrt{A} < 1$ in the integrand for 
$N_p$, then 
\begin{equation}
    m N_p < 2 (m/\Omega) M \ . \label{3d.4}
\end{equation}
Now, a negative binding energy state is defined by the condition $M - 
m N_p < 0$.  Using the inequality in Eq. (\ref{3d.4}) we see $M < m 
N_p < 2 (m/\Omega) M$, and hence the result in Eq. (\ref{3d.1}) 
follows.

Actually, the preceeding discussion does not give the tightest limit on 
the ratio $\Omega/m$.  An even tighter constraint on $\Omega/m$ can be 
found by looking at the asymptotic form for $\Phi$ as $r$ gets very 
large.  In this limit, the field equation for $\Phi$ becomes (assuming 
also that $\Phi^3$ is negligible compared to $\Phi$)
\begin{equation}
    \Phi^{\prime \prime} - \left(m^2 - \Omega^2\right) \Phi \approx 0 
                 \ , \label{3d.5}
\end{equation} 
the solution of which is $\Phi \sim {\rm exp}\left(- \sqrt{m^2 - 
\Omega^2}~r\right)$.  Thus, the stronger constraint is that $\Omega/m < 
1$.

\subsection{Rescaled equations} \label{sec3e}  
 
We will take advantage of scale-invariances of the field equations to 
redefine some of the fields, parameters, and the radial and time 
coordinates:
\begin{equation}
   x = mr \ , \ \sqrt{4 \pi G_{*}} \Phi \to \Phi \ , \ m N/\Omega \to N 
       \ , \ \Lambda/4 \pi G_{*} m^2 \to \Lambda \ , \ \Omega t/m \to t 
       \ . \label{3e.1}
\end{equation}
The field equations become (${}^{\prime} = {\rm d}/{\rm d}x$)
\begin{equation}
   {A^{\prime} \over A} = {1 - A \over x} + 2 x A \left(w + u + v
               \right) \ , \label{3e.2a}
\end{equation}
\begin{equation}
   {2 N^{\prime} \over N} = - {1 - A \over x} + 2 x A \left(w - u + v
               \right) \ , \label{3e.2b}
\end{equation}
\begin{equation}
   \left({x^2 N \varphi^{\prime} \over \sqrt{A}}\right)^{\prime} = 
         x^2 N \sqrt{A} e^{2 a(\varphi)} \alpha \left(\left[2 e^{2 
         a(\varphi)} - N^{- 2}\right] \Phi^2 + {1 \over A} 
         \left[\Phi^{\prime}\right]^2 + \lambda e^{2 a(\varphi)} 
         \Phi^4\right) \ , \label{3e.2c}
\end{equation}
\begin{equation}
   \left({x^2 N e^{2 a(\varphi)} \Phi^{\prime} \over \sqrt{A}}
         \right)^{\prime} = x^2 N \sqrt{A} e^{2 a(\varphi)} 
         \left(e^{2 a(\varphi)} - N^{- 2}\right) \Phi + \Lambda x^2 
         N \sqrt{A} e^{4 a(\varphi)} \Phi^3 \ , \label{3e.2d}
\end{equation}
where now
\begin{equation}
    w = {1 \over 2 N^2} e^{2 a(\varphi)} \Phi^2 \ , \label{3e.3a}
\end{equation}
\begin{equation}
    u = {1 \over 2} \left(1 + {\Lambda \over 2} \Phi^2\right) e^{4 
        a(\varphi)} \Phi^2 \ , \label{3e.3b}
\end{equation}
and
\begin{equation}
    v = {1 \over 2 A} \left(e^{2 a(\varphi)} \left[\Phi^{\prime}
        \right]^2 + \left[\varphi^{\prime}\right]^2\right) \ . 
        \label{3e.3c}
\end{equation}

The rescaling does not change the asymptotic value of $A$---it still 
becomes one at spatial infinity---but it does change that of $N$, 
which is now 
\begin{equation}
    \lim_{x \to \infty} N(x) = m/\Omega \ . \label{3e.4}
\end{equation}
Thus boson stars with negative binding energies will have asymptotic 
values for (rescaled) $N$ that will never be lower than $1$.  The 
total mass and total particle number are also changed to
\begin{equation}
   M = {1 \over G_{*} m} \int_0^{\infty} {\rm d}x x^2 (w + u + v) 
       \sim {1 \over G_{*} m}  \label{3e.5a}
\end{equation}
and
\begin{equation}
   N_p = {2 \over G_{*} m^2} \int_0^{\infty} {\rm d}x x^2 N \sqrt{A} 
         w \sim {1 \over G_{*} m^2}  \ . \label{3e.5b}
\end{equation}
The integrands, as well as the integrals themselves, are dimensionless.  
Hence, it is the factors in front that determine the typical values for 
$M$ and $N_p$, and they are those discussed in the Introduction.

There is one more rescaling that can be done that has no analog in 
General Relativity, and that is an invariance of the field equations 
if an arbitrary constant is added to the scalar-tensor coupling.  If we 
simultaneously do the rescaling
\begin{equation}
   e^c x \to x \ , \ e^c \Phi \to \Phi \ , \ e^c N \to N \ , \
         e^c \Lambda \to \Lambda \  \label{3e.6}
\end{equation}
on the variables defined by Eq. (\ref{3e.1}) and let $a(\varphi) + c 
\to a(\varphi)$, then the field equations remain unchanged.  It is for 
this reason that we can maintain all generality and still have boundary 
condtions for the universe today such that $a(\varphi_{\infty}) = 0$.  

\subsection{Numerical technique} \label{sec3f}

We have extended the code originally developed by Seidel and Suen 
\cite{SS90} for General Relativity.  It is based on a fourth-order 
Runge-Kutta algorithm and now solves the Scalar-Tensor Gravity 
differential equations (\ref{3e.2a})-(\ref{3e.2d}).  As mentioned 
earlier, our system requires a two parameter search to find a solution 
that satisfies the boundary conditions for $\Phi_{\infty}$ and 
$\varphi_{\infty}$ that were given earlier in Sec. \ref{sec3c}.  
Operationally, we choose a central value of the scalar field $\Phi_c$ 
first together with a guessed central value of the gravitational scalar 
field $\varphi(0)$, and integrate out to large radii for different 
values of $N(0)$.  We then check if the resulting $\varphi_{\infty}$ 
is close to our expected boundary value.  In order to judge the 
convergence of the matter scalar field $\Phi_{\infty}$, we set the 
tolerance to $10^{- 10}$, which means an asymptotic value for $\Phi$ is 
convergent if it is less than this tolerance.  

As readers will find later in Figs. 1-4, the gravitational scalar field 
falls off more slowly than the matter scalar field.  
The field equations (\ref{3e.2a})-(\ref{3e.2d}) imply that the
asymptotic behavior is $\varphi \sim \varphi_\infty + C / x$, where
$C$ is a constant. 
Therefore, at the 
numerical boundary, say $x = x_{end}$, we set the expected boundary 
values for $\varphi(x_{end})$ as
\begin{equation}
    \varphi(x_{end}) = \varphi_{\infty} + {C \over x_{end}} \ , 
                        \label{3f.1}
\end{equation}     
where
\begin{equation}
    C = - x_{end}^2 \left.{{\rm d}\varphi \over {\rm d} x}
        \right|_{x = x_{end}} \ . \label{3f.2}
\end{equation}
If the computed $\varphi(x_{end})$ is not the expected value, then we 
change $\varphi(0)$ and repeat the whole procedure.  We set the 
tolerance to judge convergence in $\varphi_{\infty}$ as $5 \times 
10^{-6}$.

The key distinctions between the matter scalar field with no nodes, and 
one with nodes, is the zero-node field only has an extremum at the 
origin, and approaches zero asymptotically, whereas all higher-node 
solutions have as many new extrema as zeroes in the matter scalar field.  
Hence, the algorithm that constructs higher-node fields searches not 
only for zeroes, but also for local extrema.  Convergence towards a 
solution satisfying the boundary conditions is effected by applying the 
same two tolerances given above.  In particular, when we search for a 
solution having $n$-nodes, then we impose the tolerance after the matter 
scalar field has passed through the $(n + 1)$-th local minimum or 
maximum.  

We have checked our code in three different ways: (i) Produced solutions 
in the large coupling limit and verified that they are identical with 
original General Relativity solutions; (ii) Replaced $a(\varphi)$ with 
$a(\varphi) + c$ and confirmed, for a given set of boundary condtions, 
that we obtain the same sequence of results; and (iii) Calculated the 
total mass $M$ three different ways (using eqs. (\ref{3c.1}-\ref{3c.3})) 
and obtained consistent results.

\section{The Equilibrium Configurations} \label{sec4}

The main goal here is to see if significant Scalar-Tensor Gravity 
effects can be generated in equilibrium state boson stars.  We strive 
for this goal in three ways: First, specific equilibrium configurations 
are constructed for a representative choice of the central value of the 
matter scalar field.  We produce different configurations by increasing 
the node-number in the matter scalar field, increasing the strength of 
the matter scalar field self-interaction, or both.  Second, we produce 
three sequences of zero-node configurations, where an individual member 
of a sequence is specified by the central value of the matter scalar 
field.  The sequences themselves are distinguished by the strength of 
the matter scalar field self-interaction.  Third, we repeat the previous 
two steps for the two different Scalar-Tensor couplings given in Eqs. 
(\ref{2.11}-\ref{2.12}) as well as for General Relativity.  
%We illustrate all of our final results in figures and tables.  

We have in Figs. 1-3 typical plots of the radial dependence of the 
matter and gravitational fields for the Brans-Dicke coupling.  Figs. 
1a, 1b, and 1c contain plots of $\Phi$ versus $x$ for $n = 0,1,2$, 
where $n$ represents the number of nodes in the matter scalar field, 
and $\Lambda = 0, 10, 100$.  The corresponding plots for $N$ and $A$ 
are given in Figs. 2a, 2b, and 2c and those for $\varphi$ in Figs. 3a, 
3b, and 3c.  All the configurations are for $\Phi_c = 0.15$, which will 
be seen in the next section to correspond to stable zero-node states, 
and $\omega_{BD} = 600$, which can be seen from Eq. (\ref{3b.2}) to be 
consistent with Solar System constraints.  The same set of plots for 
boson stars in General Relativity are essentially indistinguishable 
from those presented here.  

Notice that the effect of the nodes in the matter scalar field is to 
slightly flatten $N$ in a small region around the $x$ values where the 
nodes occur.  
For example, in Fig. 2c, we can see that
$A$ has two local extrema produced for each 
node in the scalar field. 
The local minimum occurs at precisely the 
value for $x$ where the node occurs.  The gravitational scalar $\varphi$ 
behaves similarly to $N$ near the nodes.  This behaviour for each of the 
fields is a reflection of what is occuring in the energy density: it 
becomes nearly zero in the small region around each node.  Thus, instead 
of the steady accumulation in mass that usually occurs as one moves 
radially outward from the center of a star, we have near each node 
essentially no accumulation.  The metric function $A(x)$, for instance, 
then behaves like its black hole counterpart and decreases as a node is 
approached (since the mass enclosed within $x$ remains nearly constant 
near each node).  After passing through a node, the mass will start to 
accumulate again so $A(x)$ will start to grow.  Similar remarks apply 
to the other fields.  

Initially, one might have expected the gravitational scalar field to 
have had some effect near the nodes.  After all, the mass enclosed 
inside a given radius depends on the quantities $w(x)$, $u(x)$, and 
$v(x)$.  They, in turn, depend on the gravitational scalar field (cf., 
Eqs. (\ref{3e.3a})-(\ref{3e.3c})).  It is because the gravitational 
scalar remains essentially constant (and that the realistic boundary 
conditions coming from the observational constraints force it to be 
small) that it has no significant effect on the accumulation of mass 
near the nodes.  This, in turn, is due to the fact that the matter 
scalar field serves as a {\it direct} source for the gravitational 
scalar field (cf., Eq. (\ref{3e.2c})).  

Using Tables \ref{table1} and \ref{table2} we can compare the masses, 
particle numbers, radii (which is defined to be the radial coordinate 
value from the star's center containing 95\% of the mass), and central 
value for the lapse function $N(0)$ between General Relativity and 
Brans-Dicke Gravity.  Regardless of the node-number or value for 
$\Lambda$, there are essentially no differences between the two 
theories.  If we take even larger values for $\Lambda$ then the 
differences are even smaller.  However, if we take smaller values for 
$\omega_{BD}$ then the differences get bigger (see Ref. \cite{GJ} for 
complete details on the $n = 0$ case for $\omega_{BD} = 6$).  This 
increase is natural since it is well-known that smaller values of 
$\omega_{BD}$ generally result in larger differences with General 
Relativity.  It is thus not too surprising that when $\omega_{BD} = 
600$ we cannot generate significant Scalar-Tensor effects.    

On the other hand, the quadratic coupling depends explicitly on the 
gravitational scalar field.  Hence, even if the coupling is made to 
satisfy observational constraints at spatial infinity, it is still 
possible that the gravitational scalar can be made large enough inside 
a boson star that significant deviations from General Relativity can be 
produced.  

We have constructed configurations for the quadratic coupling case with 
$\kappa = 0.38$, which is very close to the limiting value of $\kappa = 
3/8$ for this theory.  Again, we take $\Phi_c = 0.15$.  The quantitative 
results for $\Phi$, $N$ and $A$ are virtually indistinguishable from 
the previous Brans-Dicke case, and thus General Relativity.  The plot of 
the gravitational scalar $\varphi$, however, is different (see Figs. 4a, 
4b, and 4c) in that it is positive, but still maintains the same basic 
shape as the Brans-Dicke case.  The masses, particle numbers, radii, 
and central value for the lapse function $N(0)$ for zero-, one-, and 
two-node solutions for $\Phi_c = 0.15$ and $\Lambda = 0, 10, 100$ are 
listed in Table \ref{table3}.  Comparison with the previous two tables 
shows no important differences in any of the values.  Also, if we take 
larger values of $\kappa$ the differences are even smaller than those 
for $\kappa = 0.38$.  It is because the observational constraints force 
the gravitational scalar to be small at spatial infinity that this 
coupling is unable to generate any significant differences with General 
Relativity.  That is we see in Figs. 4a-c that the field equations are 
able to change the gravitational scalar by factors of order unity, but 
not factors of, say, a hundred.

The second way in which we are to extract  differences between 
General Relativity and Scalar-Tensor Gravity is to consider sequences 
of zero-node configurations, where individual members in a sequence 
are delimited by their value for $\Phi_c$.  In Fig. 5 we have a plot 
of $M,N_p$ vs $\Phi_c$, for $\Lambda = 0, 10, 100$.  For each value of 
$\Lambda$ we see that there is an absolute maximum mass, and number of 
particles, that any boson star can have, exactly as in General 
Relativity.  (Similar results for the mass were obtained by Gunderson 
and Jensen \cite{GJ} for the Brans-Dicke coupling with $\omega_{BD} = 
6$, as well as the couplings considered by Torres \cite{torres}.)  Not 
unexpected, the value for the maximum mass and particle number are not 
changed significantly from the General Relativity results.  In Fig. 6 
we have plots of the binding energy versus the particle number.  We 
note that all branches starting from $\Phi_c = 0$ have negative binding 
energies, just like General Relativity.  The corresponding plots for 
the quadratic coupling (for $p = 10$, and $\kappa = 0.38$) contain 
essentially the same features as Brans-Dicke Gravity.

So, we see that the observational constraints are so restrictive that 
no significant deviations from General Relativity can be produced.  
However, we can at least extract some generic behaviour pertinent to 
the gravitational scalar, and for boundary conditions consistent with 
observational constraints,  from  Figs. 7 and 8.  They  are 
the Brans-Dicke and quadratic coupling graphs, respectively, of $M$ 
vs $\varphi(0)$ for a sequence of zero-node configurations having 
$\Lambda = 0, 10, 100$.  Each curve is parameterized by $\Phi_c$, and 
starts at $M = 0$.  As $\Phi_c$ gets bigger, then $\varphi(0)$ decreases 
and $M$ increases.  At precisely the same value of $\Phi_c$ where the 
maximum mass occurs, $M$ reaches its maximum value in Figs. 7 and 8; 
also, the local minimum $M$ in both Figs. 7 and 8 corresponds to the first 
local minimum in Fig. 5.  

The first point to be grasped from these plots is that the central value 
of the gravitational scalar has an absolute minimum value.  The second 
is that there is a region of the graph where the same value for 
$\varphi(0)$ corresponds to two different mass values.  There is even 
the ``crossover'' point where two configurations have exactly the same 
mass.  (It can be inferred from the discussion below that one 
configuration is stable, and the other is not.)  The final point is 
that there is the asymptotic region where the mass $M$ oscillates 
slightly in value, but $\varphi(0)$ continues to grow.  This is certainly 
surprising, since one would expect a growing $\varphi(0)$ (which is 
related to the Newton gravitational coupling between two particles) to 
force the mass to change monotonically. 

\section{Stability Analysis via Catastrophe Theory} \label{sec5}

Catastrophe theory is a relatively new mathematical tool to explain 
a variety of changes of state in physical systems \cite{VA}.  Rigorous 
theorems have been established demonstrating its validity.  It is 
particularly adept at extracting discontinuous changes of state when 
gradual changes occur in the system parameters.  It has been used in 
General Relativity to analyze the stability of black holes \cite{KOK}, 
non-abelian black holes \cite{MTTM}, and boson stars \cite{cata}.  
Most importantly for the present work, is that application of 
catastrophe theory to boson stars \cite{cata} 
delivers the same results for changes 
in stability that are obtained from dynamical, numerical analysis 
\cite{SS90}.

Catastrophe theory can be used to determine stability if the system 
under analysis develops bifurcations, or cusps, when curves existing in 
its so-called equilbrium space are projected into its lower-dimensional 
control parameter space.  The variables for the equilibrium space are 
given by a {\it potential function}, {\it control parameters}, and {\it 
state variables}.  The variables for the control parameter space are the 
potential function and the control parameters.  For the problem at hand, 
the equilibrium space is three-dimensional and the control parameter 
space two-dimensional, with the binding energy $M - m N_p$ taken to be
the potential function, the total particle number $N_p$ chosen to be the 
control parameter, and the central value of the matter scalar field 
$\Phi_c$ taken to be the state variable.  An equivalent variation on 
this choice for the equilibrium space and the control parameter space 
is to replace the total particle number with the mass. 

The appearance of a cusp in a curve's projection is not enough to 
determine a change in stability.  Rather, if it is known that a system 
is stable along one branch of a curve in the control parameter space, 
then it will become unstable on the next branch formed at a cusp.  The 
trick is in determining stability along the previous branch.  Like 
the case of General Relativity \cite{cata}, we can 
assume that boson stars along the very first branches in Fig. 6 (i.e., 
those that start at $N_p = 0$ and end at the cusp labled $B_1$) are 
stable. 
They are the only ones composed entirely of negative 
binding energy states, they contain the flat-space limit, and 
furthermore, results of a dynamical, 
numerical analysis demonstrate that they 
are stable against small perturbations \cite{BCSS}.

%Fig. 9 is the equilibrium space for the Brans-Dicke coupling for 
%$\Lambda = 100$. 
Fig. 9 is the typical equilibrium space for the Brans-Dicke coupling.
 The projection into the $(N_p,\Phi_c)$ plane at $M - 
m N_p = 0$ gives the same curve for the particle number as in Fig. 5.  
The projection into the $(N_p,M - m N_p)$ plane at the maximum 
$\Phi_c$ value gives the same curve as in Fig. 6.  It is this second 
projection that yields the control parameter space.  Hence, the cusp 
labeled $B_1$ in Fig. 6 represents a change in stability, since the 
configurations on the branch leading up to $B_1$ are all stable.  At 
the cusp we have $\Phi_c = 0.09$ and a mass of $M = 2.255/G_* m$, which 
is not significantly different from the corresponding General Relativity 
result (see \cite{cata}).  
The cusp at $B_2$ does not represent a 
change in stability since the binding energies near this point are 
positive; likewise, for the cusp at $B_3$.  
As one might expect, the 
maximum stable masses for $\Lambda = 0$ and $10$ are not significantly 
different from their General Relativity values.  Neither are the 
quadratic coupling results, since the equilibrium space and control 
parameter space are virtually identical to what is depicted in 
Fig. 9.

\section{Configurations in the Earlier Universe} \label{sec6}

We have seen that the attractor nature to General Relativity of the 
quadratic model extends also to a sequence of equilibrium boson star 
configurations.  Therefore, we repeat here the same calculations as 
the previous section but for earlier eras for the universe according 
to the quadratic coupling model.  Going to earlier eras means taking 
smaller values for $p$ (which, recall, is related to the redshift via 
$z \sim e^{10-p}-1$) in Eq. (\ref{3b.1}).  Our interest will be 
focused on the maximum allowed stable mass.

A not so trivial point is that all our calculations must be made 
within the {\it same} Scalar-Tensor theory.  What this means is that 
we are allowed to choose the additive constant to the coupling function 
$a(\varphi)$ only once, and not change it when we change the value for 
$p$.  We will stick with the choice that makes $a(\varphi_{\infty}) = 
0$ at $p = 10$.  Thus, $a(\varphi_{\infty})$ is not zero for any other 
value of $p$.  This complicates the determination of the ADM mass, 
since for $p \neq 10$ it can no longer be obtained from Eq. 
(\ref{3e.5a}).  Fortunately, we can use the energy functional of Eq. 
(\ref{3a.7}) to determine the total mass, once the rescaling relations 
of Eqs. (\ref{3e.1}) and (\ref{3e.6}) are properly taken into account.

In Fig. 10 we have a plot of $M$ vs $\Phi_c$.  We only go far enough 
so that the first extremum can be clearly identified, which, as we 
know from the previous section, corresponds to the maximum stable mass.  
Clearly, the main effect is to decrease the mass, as suggested by Torres 
\cite{torres}.  If we go to early enough cosmological eras ($p < 5$), 
however, we find that the binding energy is positive, for all 
values of $\Phi_c$ (see Fig. 11).  Furthermore, the cusp that appears 
at $B_1$ in Fig. 6 is clearly gone for $p < 5$.  This indicates, then,
that no stable Scalar-Tensor Gravity boson star can be formed during 
early enough cosmological times. 

\section{Concluding Remarks} \label{sec7}

The main point of this work is to see if significant Scalar-Tensor 
Gravity effects can be generated in equilibrium state boson stars, 
even when the fairly restrictive observational constraints are imposed.  
In contrast with earlier works \cite{GJ,torres}, we constructed not only 
zero-node solutions, but also one- and two-node solutions.  We also 
illustrated some of the interesting features of the gravitational
scalar for the zero-node case, and used catastrophe theory to 
determine the maximum stable mass for the zero-node states.  Finally, we 
confirmed Torres' \cite{torres} suggestion that masses decrease below 
corresponding General Relativity masses for boson stars formed in 
earlier eras of the history of the universe.  

However, our results also show that stars formed too early are unstable.  
This suggests, for instance, that a scenario whereby boson stars are 
formed in some cosmological phase transition will not work in 
Scalar-Tensor Gravity.  Moreover, there is every likelihood that {\it 
any} type of stellar object will suffer from the same instability.  For 
instance, Harada \cite{TH} has recently shown that there are a range of 
values for the coupling for which perfect fluid stars are unstable, 
although no explicit connection with the ambient cosmological conditions 
is made.

We did not apply catastrophe theory to the higher-node solutions, because 
we know, from numerical evolutions \cite{BSS}, that they are all unstable 
in General Relativity.  Dynamical 1-D evolutions are being performed to 
see if the higher-node solutions are generically unstable in Scalar-Tensor 
Gravity.  As mentioned earlier, the code developed for these evolutions 
confirms that zero-node configurations on the initial branch of Fig. 6 are 
stable.  It is also being investigated whether the zero-node states on the 
other branches collapse, or disperse.  So far, the results do 
substantiate the conclusions obtained here using catastrophe theory.  All 
these results are currently under preparation \cite{BCSS}. 

\acknowledgements
We thank Ed Seidel and Wai-Mo Suen for letting us modify their General 
Relativity boson star code \cite{SS90} to Scalar-Tensor gravity, and 
Jay Balakrishna for pointing out subtleties in the higher-node algorithm.  
We also thank Wai-Mo Suen and Franz Schunck for useful conversations.  H. 
S. was partially supported by the grant NSF PHYS 96-00507, 96-00049, and 
NASA NCCS 5-153.

\vfill

%----------------------------------------------------------- tables --
\eject
\begin{table}
\begin{tabular}{ccccccc} 
  $n$ & $\Lambda$ & $M$ & $N_p$ & $R$ & $N(0)$ & $m/\Omega$ \\ \hline \hline   
   $~$ & 0 & 0.5916 & 0.6056 & 8.760 & 0.8255 & 0.9210\\
   0 & 10 & 0.8573 & 0.8904 & 8.855 & 0.7628 & 0.8999\\
   $~$ & 100 & 1.963 & 2.034 & 9.485 & 0.4655 & 0.8575\\ \hline
   $~$ & 0 & 1.288 & 1.316 & 18.34 & 0.8181 & 0.9482\\ 
   1 & 10 & 1.512 & 1.554 & 17.27 & 0.7723 & 0.9353\\ 
   $~$ & 100 & 2.531 & 2.612 & 14.09 & 0.4820 & 0.8881\\ \hline
   $~$ & 0 & 1.983 & 2.025 & 28.26 & 0.8170 & 0.9636\\ 
   2 & 10 & 2.198 & 2.254 & 26.21 & 0.7756 & 0.9528\\ 
   $~$ & 100 & 3.112 & 3.199 & 19.20 & 0.4887 & 0.9007\\ 
\end{tabular}
\bigskip
\noindent
\caption{Listed are some sample values for General Relativity, for
         ADM mass (modulo $1/ G_{*} m$), particle number (modulo $1/G_{*} 
         m^2$), radius (modulo $1/m$), central value for $N/(m/\Omega)$, 
         and $m/\Omega$ for $\Phi_c = 0.15$.} \label{table1}
\end{table}

\begin{table}
\begin{tabular}{ccccccc} 
  $n$ & $\Lambda$ & $M$ & $N_p$ & $R$ & $N(0)$ & $m/\Omega$ \\ \hline \hline   
   $~$ & 0 & 0.5914 & 0.6021 & 8.750 & 0.8289 & 0.9247\\
   0 & 10 & 0.8571 & 0.8857 & 8.850 & 0.7679 & 0.9057\\
   $~$ & 100 & 1.964 & 2.026 & 9.475 & 0.4759 & 0.8765\\ \hline
   $~$ & 0 & 1.288 & 1.312 & 18.32 & 0.8131 & 0.9424\\ 
   1 & 10 & 1.512 & 1.550 & 17.28 & 0.7671 & 0.9289\\ 
   $~$ & 100 & 2.531 & 2.604 & 14.08 & 0.4843 & 0.8921\\ \hline
   $~$ & 0 & 1.983 & 2.020 & 28.25 & 0.8110 & 0.9564\\ 
   2 & 10 & 2.197 & 2.249 & 26.20 & 0.7686 & 0.9441\\ 
   $~$ & 100 & 3.113 & 3.192 & 19.20 & 0.4912 & 0.9051\\ 
\end{tabular}
\bigskip
\noindent
\caption{Listed are some sample values for Brans-Dicke, for 
         ADM mass (modulo $1/ G_{*} m$), particle number  
         (modulo $1/ G_{*} m^2$), radius (modulo $1/m$), central 
         value for $N/(m/\Omega)$, and $m/\Omega$ for $\omega_{BD} = 600$ 
         and $\Phi_c = 0.15$.}
         \label{table2}
\end{table}

\begin{table}
\begin{tabular}{ccccccc} 
  $n$ & $\Lambda$ & $M$ & $N_{p}$ & $R$ & $N(0)$ & $m/\Omega$ \\ \hline \hline   
   $~$ & 0 & 0.5916 & 0.6021 & 8.800 & 0.8271 & 0.9228\\
   0 & 10 & 0.8573 & 0.8856 & 8.875 & 0.7653 & 0.9029\\
   $~$ & 100 & 1.963 & 2.025 & 9.475 & 0.4681 & 0.8624\\ \hline
   $~$ & 0 & 1.288 & 1.312 & 18.38 & 0.8128 & 0.9427\\ 
   1 & 10 & 1.512 & 1.550 & 17.30 & 0.7671 & 0.9290\\ 
   $~$ & 100 & 2.531 & 2.602 & 14.08 & 0.4842 & 0.8922\\ \hline
   $~$ & 0 & 1.983 & 2.020 & 28.27 & 0.8068 & 0.9517\\ 
   2 & 10 & 2.198 & 2.245 & 26.22 & 0.7682 & 0.9438\\ 
   $~$ & 100 & 3.112 & 3.112 & 19.20 & 0.4861 & 0.8960\\ 
\end{tabular}
\bigskip
\noindent
\caption{Listed are some sample values for the quadratic coupling, for
         ADM mass (modulo $1/ G_{*} m$), particle number (modulo $1/ G_{*} 
         m^2$), radius (modulo $1/m$), central value for $N/(m/\Omega)$, 
         and $m/\Omega$ for $\kappa = 0.38$ and $\Phi_c = 0.15$.} \label{table3}
\end{table}

%-------------------------------------------------- figure captions --
\newpage

{\bf Figure captions} \\ ~ \\

\begin{figure}[h]
%\vspace*{1.5cm}
\setlength{\unitlength}{1in}
\begin{picture}(7.5,2.5)
\put(-0.0,-2.5){\epsfxsize=7.0in \epsffile{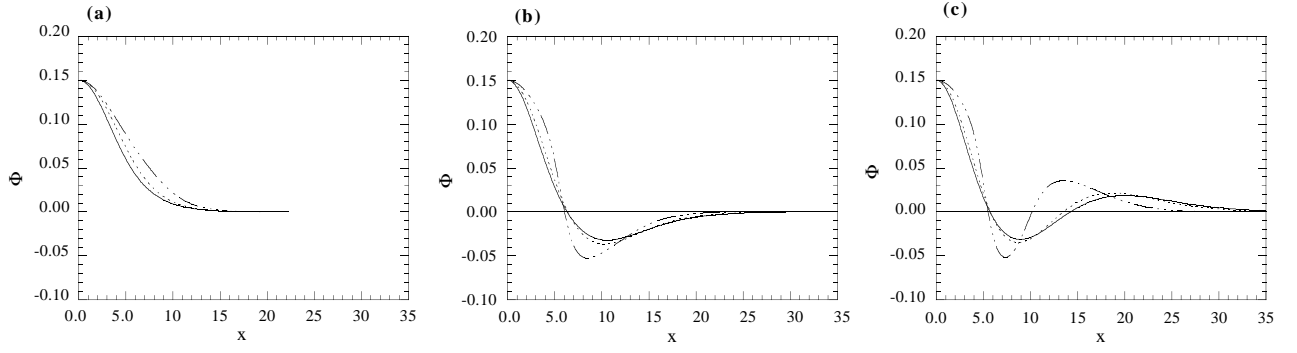} }
\end{picture}
\caption[shinka01]{
Sample configurations of equilibrium state boson stars in 
the Brans-Dicke  theory. 
Matter scalar field  $\Phi$ is plotted  in the case of 
 node=1 (a), 2 (b), and 3(c), respectively. Solid, dotted and 
three-dot-dash lines are for $\Lambda=0,10$ and 100 case, 
respectively.
}
\label{shinka01}
\end{figure}

\noindent
{\bf Fig.\ref{shinka01}}\\
\noindent
Sample configurations of equilibrium state boson stars in 
the Brans-Dicke  theory. 
Matter scalar field  $\Phi$ is plotted  in the case of 
 node=1 (a), 2 (b), and 3(c), respectively. Solid, dotted and 
three-dot-dash lines are for $\Lambda=0,10$ and 100 case, 
respectively.

~

\begin{figure}[h]
%\vspace*{1.5cm}
\setlength{\unitlength}{1in}
\begin{picture}(7.5,2.5)
\put(-0.0,-2.5){\epsfxsize=7.0in \epsffile{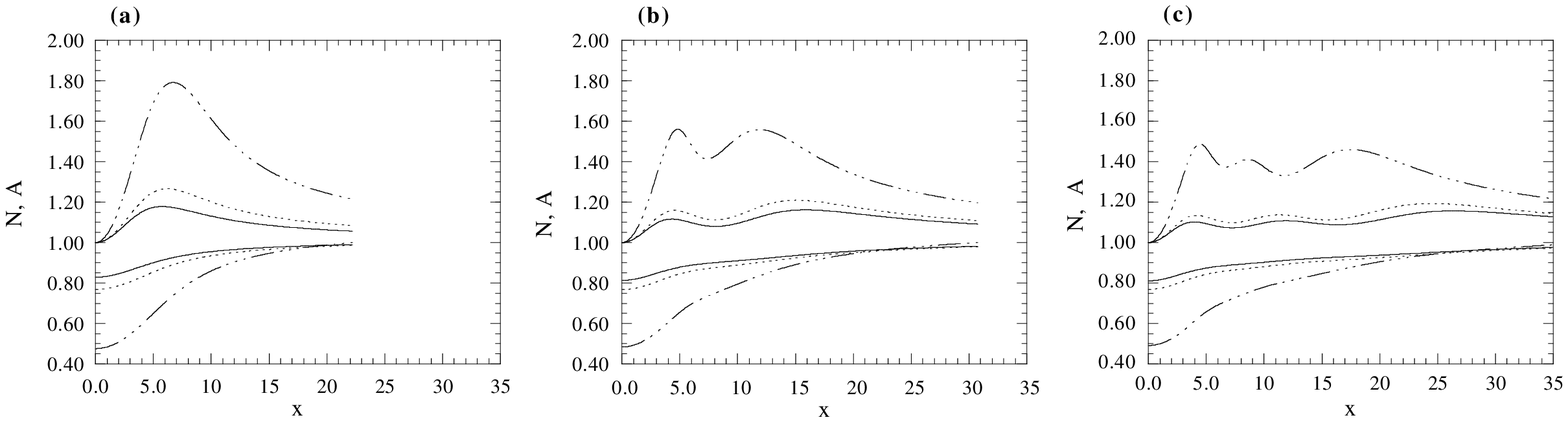} }
\end{picture}
\caption[shinka02]{
Lapse $N$ and the metric $A$ for the
solutions plotted in Fig.\ref{shinka01}.  
The lapse $N$ has been re-scaled to its original 
form so that its asymptotic value is 1.
}
\label{shinka02}
\end{figure}

\noindent
{\bf Fig.\ref{shinka02}}\\
\noindent
Lapse $N$ and the metric $A$ for the solutions plotted in 
Fig.\ref{shinka01}.  The lapse $N$ has been re-scaled to its original 
form so that its asymptotic value is 1.  

~ 

\begin{figure}[h]
%\vspace*{1.5cm}
\setlength{\unitlength}{1in}
\begin{picture}(7.5,2.5)
\put(-0.0,-2.5){\epsfxsize=7.0in \epsffile{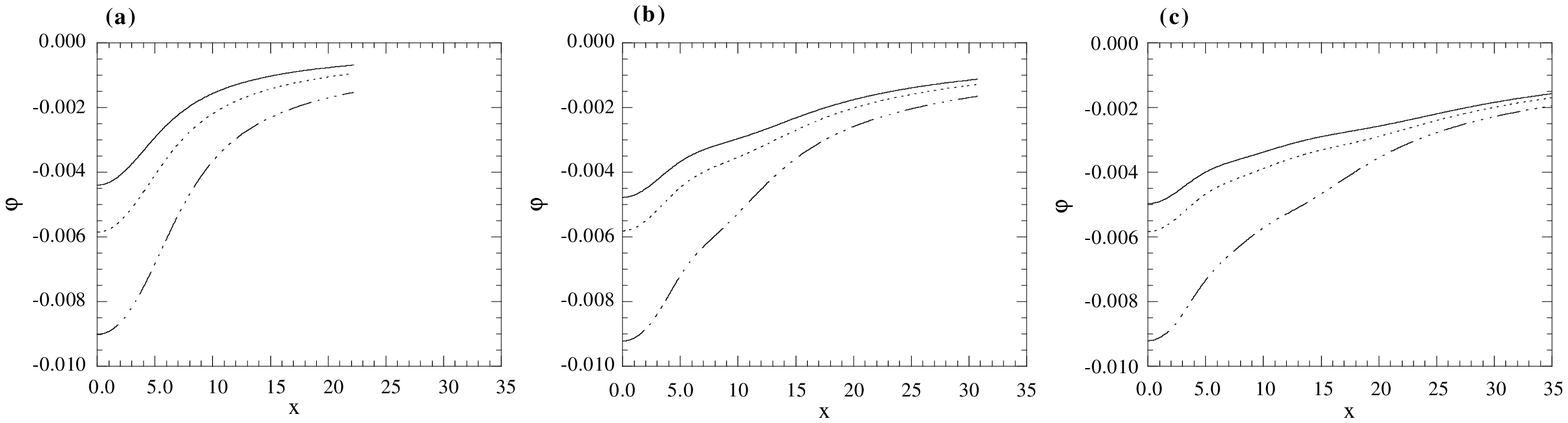} }
\end{picture}
\caption[shinka03]{
Gravitational scalar field $\varphi$ for the
solutions plotted in Fig.\ref{shinka01}.
}
\label{shinka03}
\end{figure}

\noindent
{\bf Fig.\ref{shinka03}}\\
\noindent
Gravitational scalar field $\varphi$ for the
solutions plotted in Fig.\ref{shinka01}.

~

\begin{figure}[h]
%\vspace*{1.5cm}
\setlength{\unitlength}{1in}
\begin{picture}(7.5,2.5)
\put(-0.0,-2.5){\epsfxsize=7.0in \epsffile{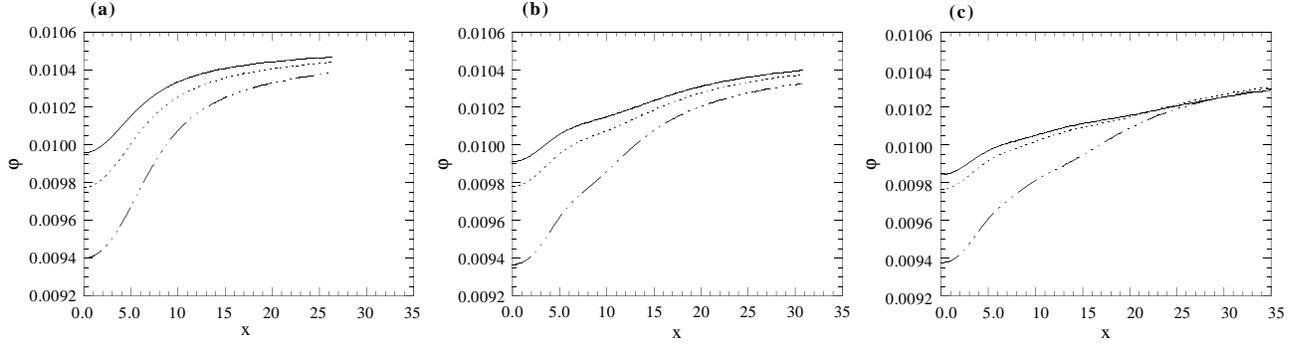} }
\end{picture}
\caption[shinka04]{
Gravitational scalar field $\varphi$ for the
quadratic coupling model. The value for $\Phi_c$ is
the same as in Fig.\ref{shinka01}.
}
\label{shinka04}
\end{figure}

\noindent
{\bf Fig.\ref{shinka04}}\\
\noindent
Gravitational scalar field $\varphi$ for the
quadratic coupling model. The value for $\Phi_c$ is
the same as in Fig.\ref{shinka01}.

~

\begin{figure}[h]
%\vspace*{1.5cm}
\setlength{\unitlength}{1in}
\begin{picture}(7.5,4.0)
\put(.5 , 0.0){\epsfxsize=5.0in \epsffile{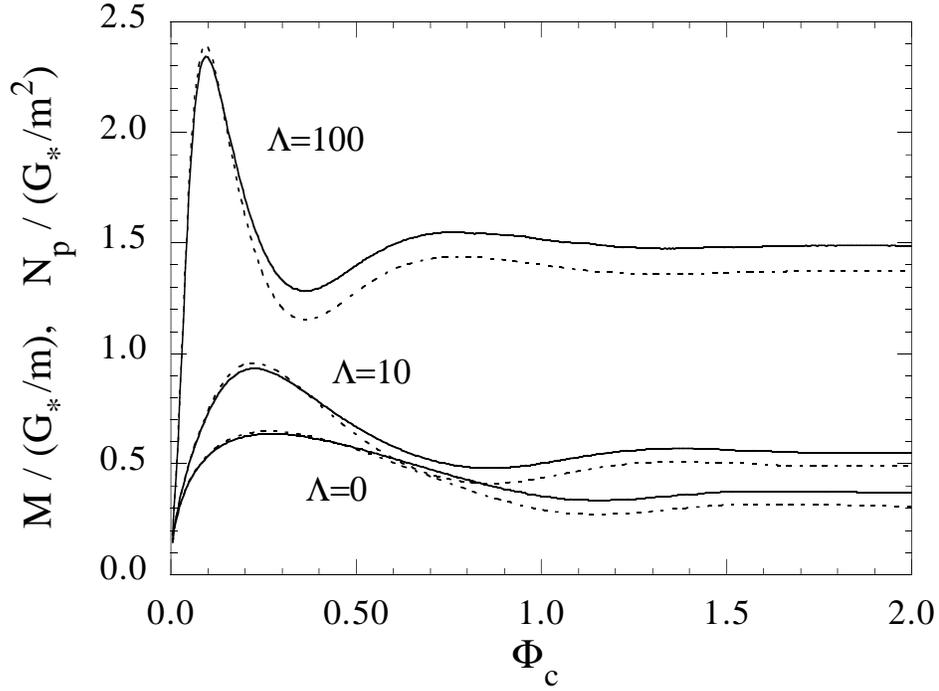} }
\end{picture}
\caption[shinkai05]{
Mass and particle numbers versus central matter scalar field
in the Brans-Dicke theory.
}
\label{shinka05}
\end{figure}

\noindent
{\bf Fig.\ref{shinka05}}\\
\noindent
Mass and particle numbers versus central matter scalar field
in the Brans-Dicke theory.

~

\begin{figure}[h]
%\vspace*{1.5cm}
\setlength{\unitlength}{1in}
\begin{picture}(7.5,4.0)
\put(.5 , 0.0){\epsfxsize=5.0in \epsffile{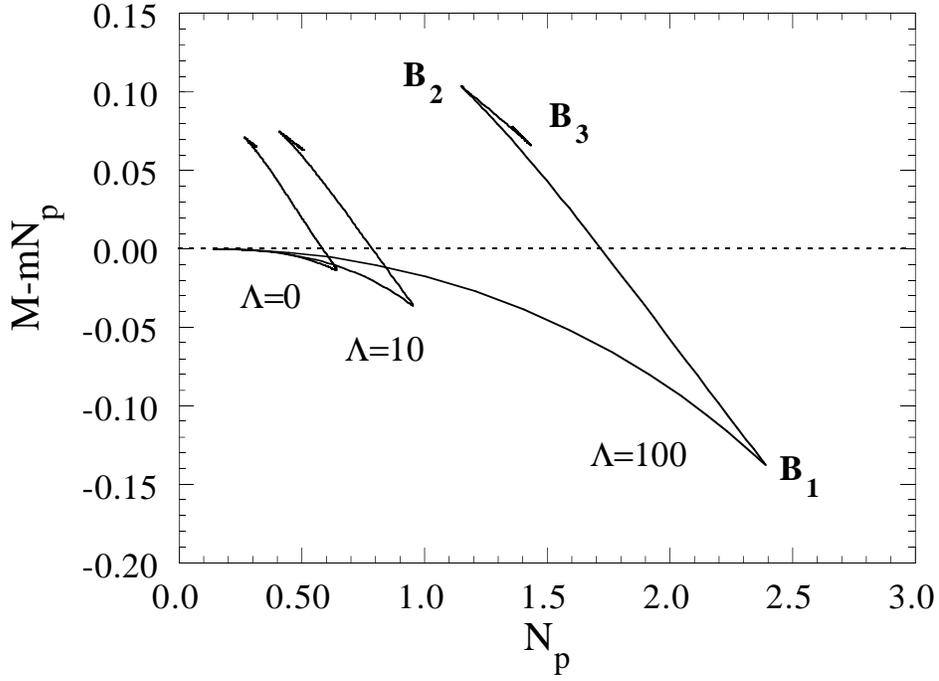} }
\end{picture}
\caption[shinka06]{
Binding energy $M-mN_p$ versus particle number $N_p$ 
in the quadratic coupling model. 
}
\label{shinka06}
\end{figure}

\noindent
{\bf Fig.\ref{shinka06}}\\
\noindent
Binding energy $M-mN_p$ versus particle number $N_p$ 
in the quadratic coupling model. 

~

\begin{figure}[h]
%\vspace*{1.5cm}
\setlength{\unitlength}{1in}
\begin{picture}(7.5,4.0)
\put(.5 , 0.0){\epsfxsize=5.0in \epsffile{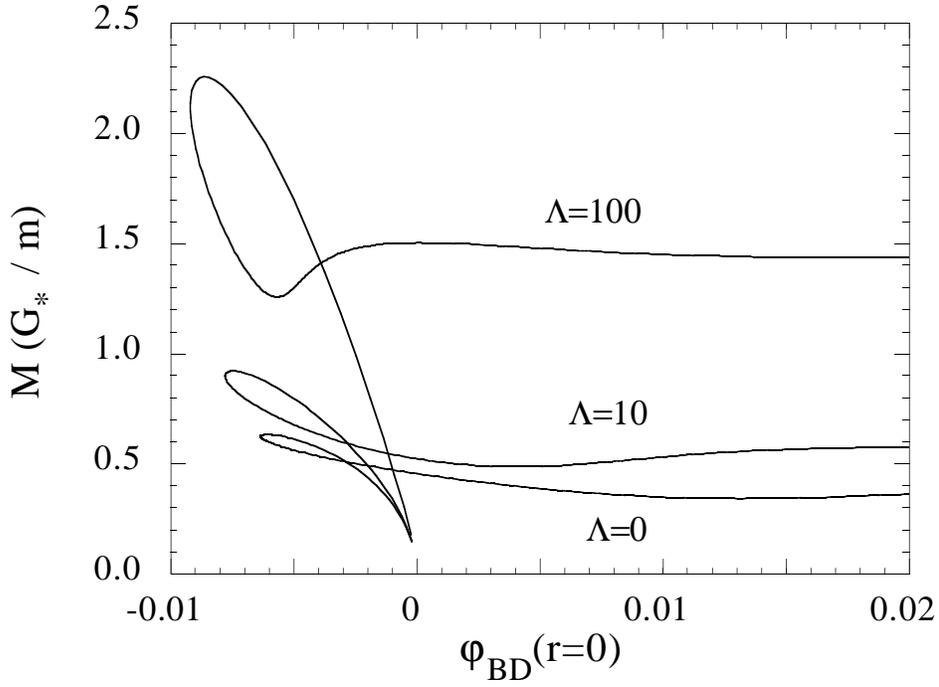} }
\end{picture}
\caption[shinka07]{
A sequence of equilibrium solutions in the Brans-Dicke theory.
Mass  versus central gravitational  scalar field value  is plotted.
}
\label{shinka07}
\end{figure}

\noindent
{\bf Fig.\ref{shinka07}}\\
\noindent
A sequence of equilibrium solutions in the Brans-Dicke theory.
Mass  versus central gravitational  scalar field value  is plotted.

~

\begin{figure}[h]
%\vspace*{1.5cm}
\setlength{\unitlength}{1in}
\begin{picture}(7.5,4.0)
\put(.5 , 0.0){\epsfxsize=5.0in \epsffile{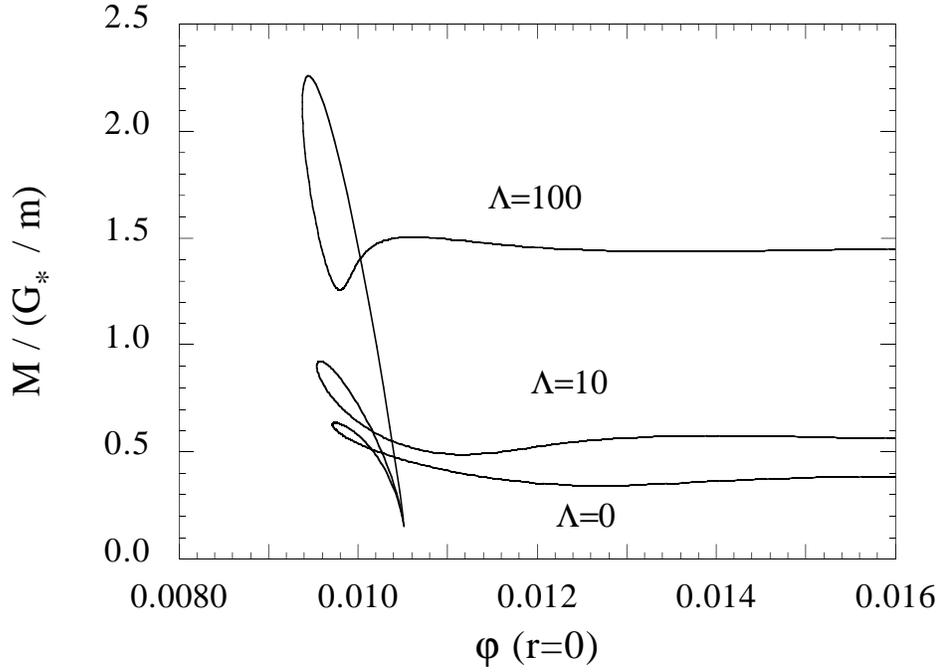} }
\end{picture}
\caption[shinka08]{
A sequence of equilibrium solutions in the quadratic coupling model.
Mass  versus central gravitational  scalar field value is plotted.
}
\label{shinka08}
\end{figure}

\noindent
{\bf Fig.\ref{shinka08}}\\
\noindent
A sequence of equilibrium solutions in the quadratic coupling model.
Mass  versus central gravitational  scalar field value is plotted.

~

\begin{figure}[h]
%\vspace*{1.5cm}
\setlength{\unitlength}{1in}
\begin{picture}(7.5,4.0)
\put(.0 , 0.5){\epsfxsize=5.5in \epsffile{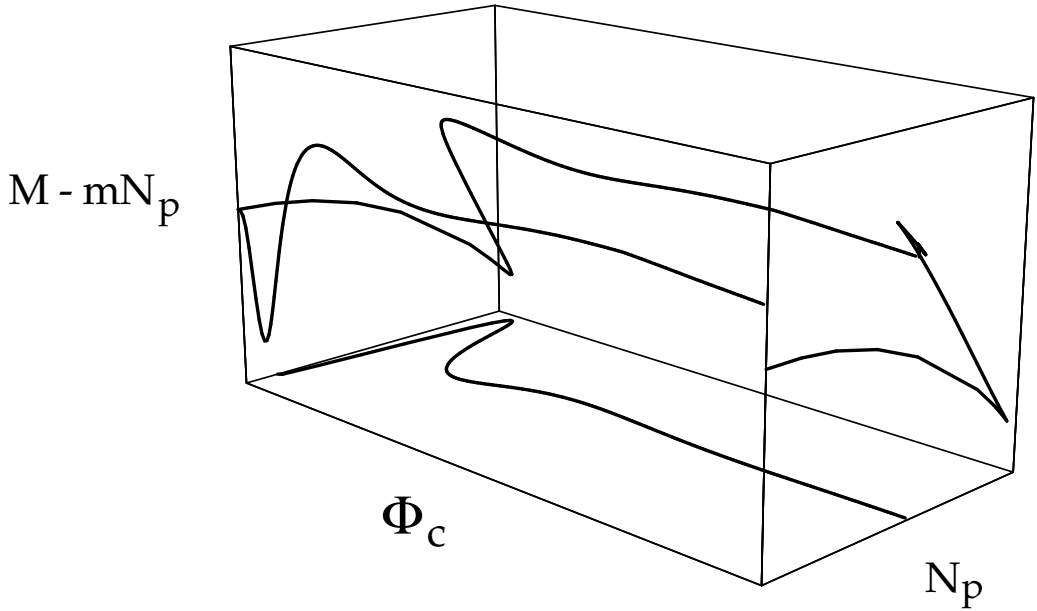} }
\end{picture}
\caption[shinka09]{
Equilibrium configurations in the equilibrium space. Binding 
energy  $M-mN_p$,
particle number $N_p$ and the center matter scalar field
$\Phi_c$ are taken as the
potential function,
control parameter, and 
state parameter, respectively.
One equilibrium sequence line and three projected lines of it onto
2-parameter planes are shown. The plot on the $(N_p, M-mN_p)$ plane is
identical with Fig.\ref{shinka06}.

}
\label{shinka09}
\end{figure}

\noindent
{\bf Fig.\ref{shinka09}}\\
\noindent
Equilibrium configurations in the equilibrium space. Binding 
energy  $M-mN_p$,
particle number $N_p$ and the center matter scalar field
$\Phi_c$ are taken as the
potential function,
control parameter, and 
state parameter, respectively.
One equilibrium sequence line and three projected lines of it onto
2-parameter planes are shown. The plot on the $(N_p, M-mN_p)$ plane is
identical with Fig.\ref{shinka06}.

~

\begin{figure}[h]
%\vspace*{1.5cm}
\setlength{\unitlength}{1in}
\begin{picture}(7.5,4.0)
\put(.5 , 0.0){\epsfxsize=5.0in \epsffile{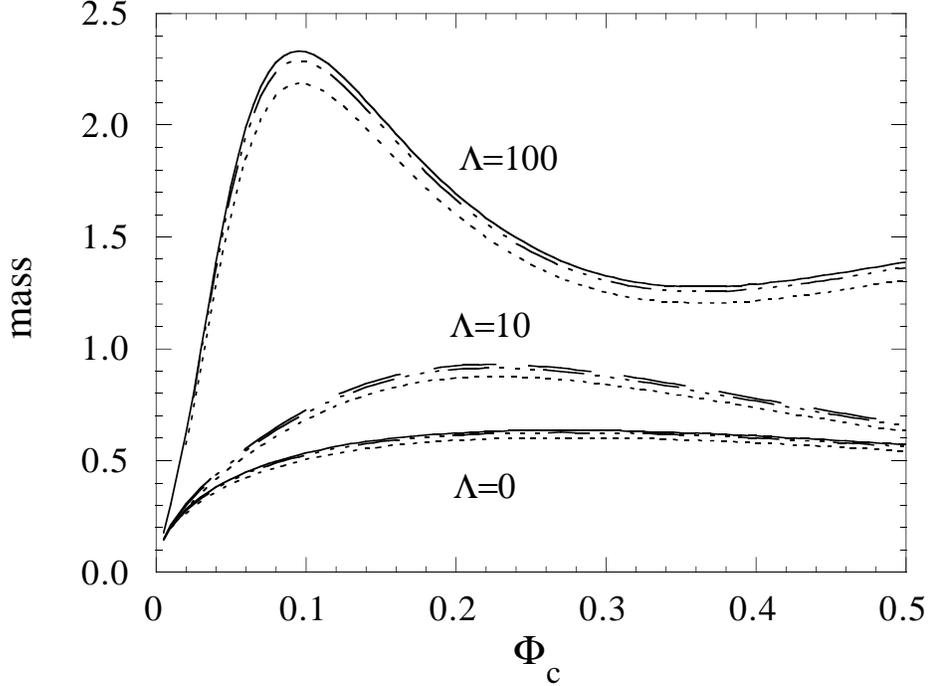} }
\end{picture}
\caption[shinka10]{
Changing the boundary condition for $\varphi_\infty$ via
the cosmological solution of the quadratic coupling.
The sequences of equilibrium solutions for cosmological time parameter 
$p=10$(solid line), 6(three-dots-dash line) and 5(dotted line) are plotted. 
The solid lines are identical with those in Fig.\ref{shinka05}.

}
\label{shinka10}
\end{figure}

\noindent
{\bf Fig.\ref{shinka10}}\\
\noindent
Changing the boundary condition for $\varphi_\infty$ via
the cosmological solution of the quadratic coupling.
The sequences of equilibrium solutions for cosmological time parameter 
$p=10$(solid line), 6(three-dots-dash line) and 5(dotted line) are plotted. 
The solid lines are identical with those in Fig.\ref{shinka05}.

~

\begin{figure}[h]
%\vspace*{1.5cm}
\setlength{\unitlength}{1in}
\begin{picture}(7.5,4.0)
\put(.5 , 0.0){\epsfxsize=5.0in \epsffile{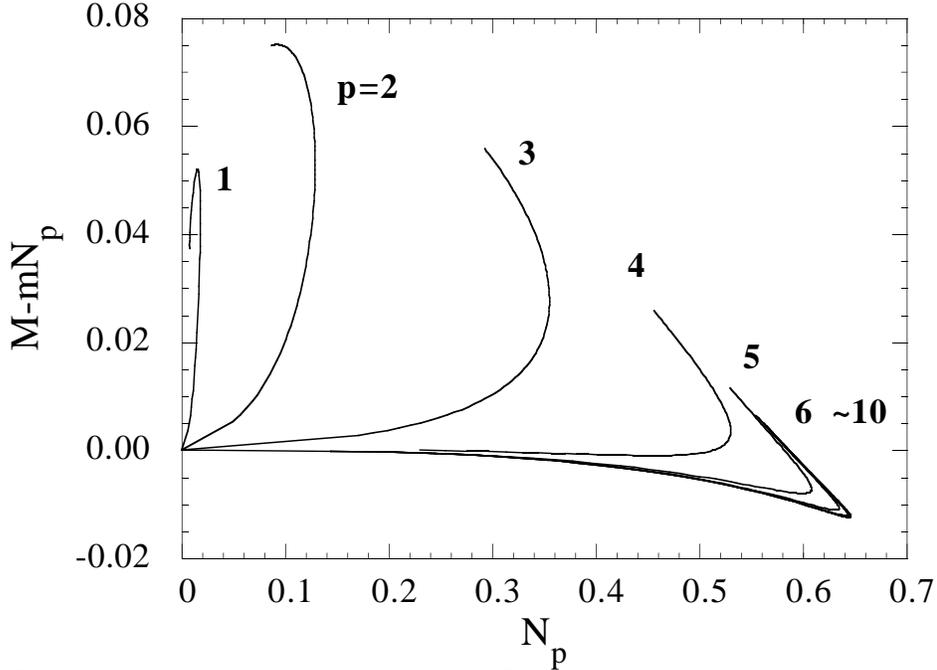} }
\end{picture}
\caption[shinka11]{
Binding energy vs. particle number $N_p$ for different 
cosmological time $p=10,9,\cdots,1$.  The case for $p=10$
is the same as Fig.\ref{shinka06}.  Only the range $0 \leq 
\Phi_c \leq 0.5$ is plotted.
 
}
\label{shinka11}
\end{figure}

\noindent
{\bf Fig.\ref{shinka11}}\\
\noindent
Binding energy vs. particle number $N_p$ for different 
cosmological time $p=10,9,\cdots,1$.  The case for $p=10$
is the same as Fig.\ref{shinka06}.  Only the range $0 \leq 
\Phi_c \leq 0.5$ is plotted.

%---------------------------------------------- figures and  captions --

\newpage

%****************************** Fig.\ref{shinka01}  >>>>>.
%******************************     \ref{shinka01}  <<<<<.
%****************************** Fig.\ref{shinka02}  >>>>>.
%******************************     \ref{shinka02}  <<<<<.

%****************************** Fig.\ref{shinka03}  >>>>>.
%******************************     \ref{shinka03}  <<<<<.

%****************************** Fig.\ref{shinka04}  >>>>>.
%******************************     \ref{shinkai04}  <<<<<.
%****************************** Fig.\ref{shinkai05}  >>>>>.
%******************************     \ref{shinka05}  <<<<<.

%****************************** Fig.\ref{shinka06}  >>>>>.
%******************************     \ref{shinka06}  <<<<<.

%****************************** Fig.\ref{shinka07}  >>>>>.
%******************************     \ref{shinka07}  <<<<<.

%****************************** Fig.\ref{shinka08}  >>>>>.
%******************************     \ref{shinka08}  <<<<<.

%****************************** Fig.\ref{shinka09}  >>>>>.
%******************************     \ref{shinka09}  <<<<<.

%****************************** Fig.\ref{shinka10}  >>>>>.
%******************************     \ref{shinka10}  <<<<<.

%****************************** Fig.\ref{shinka11}  >>>>>.
%******************************     \ref{shinka11}  <<<<<.

\end{document}